\documentclass[aps,prl,groupedaddress,twocolumn,showpacs]{revtex4}
\usepackage{graphicx,graphics,psfrag,amsmath,calc,amssymb}

\begin{document}

\title{Coexistence of Spin Density Wave and Triplet Superconductivity}
\author{Wei Zhang}
\author{C. A. R. S\'{a} de Melo}
\affiliation{School of Physics, Georgia Institute of Technology, 
Atlanta, Georgia 30332}


\begin{abstract}
We discuss the possibility of coexistence of spin density wave 
(antiferromagnetism) and triplet superconductivity 
as a particular example of a broad class 
of systems where the interplay of magnetism and superconductivity 
is important. We focus on the case of quasi-one-dimensional metals,
where it is known experimentally that antiferromagnetism
is in close proximity to triplet superconductivity in the temperature
versus pressure phase diagram. Over a narrow range of pressures, we propose 
an intermediate non-uniform phase consisting of alternating
antiferromagnetic and triplet superconducting stripes. 
Within the non-uniform phase there are also changes between two and
three dimensional behavior.

\end{abstract}

\maketitle


The competition or coexistence of magnetic order and superconductivity 
is a very important problem in condensed matter physics. There is a broad
class of systems that present magnetic order and superconductivity in 
close vicinity. One of the most important systems are 
the Copper Oxides, where singlet superconductivity is found next 
to antiferromagnetism~\cite{dagotto-94}. 
In addition, striped phases, where coexistence
of antiferromagnetic order and singlet d-wave superconductivity, were
observed in Copper Oxides~\cite{tranquada-96}. 
Another system where magnetism and superconductivity are 
intertwined is Strontium Ruthenate, where the proximity to ferromagnetism 
has been argued as being important to the existence of triplet 
superconductivity in these materials~\cite{maeno-94}.
Furthermore the newly discovered ferromagnetic 
superconductors $ZrZn_2$ and $UGe_2$ have stimulated a debate on the coexistence
of ferromagnetism and triplet or singlet superconductivity~\cite{lonzarich-01,
saxena-00}.
However, unlike any of these previous examples, we will discuss 
in this manuscript a system
which may exhibit coexistence of antiferromagnetism (AF) and 
triplet superconductivity (TS).

New experiments on 
quasi-one-dimensional superconductors in high magnetic fields have
shown that TS  is strongly affected
by the proximity to an AF phase characterized by insulating 
spin density wave (SDW) order~\cite{lee-02a}. 
From now on we will use interchangeably SDW and AF. 
Motivated by these
experiments and the known phase diagram of
quasi-one-dimensional ${\rm (TMTSF)_2 PF_6}$ under pressure we propose a new 
phase for quasi-one-dimensional systems where AF (SDW)
and TS coexist. 
The coexistence of these phases implies that the new state 
is non-uniform, with alternating stripes of 
SDW and TS, due to the appearance of a negative
interface energy between SDW and TS regions. 
As indicated in the schematic 
phase diagram (Fig.1),
the inhomogeneous intermediate phase is expected to
exist over a narrow range of pressures
$\Delta P = P_2 (T)  - P_1 (T)$ around $P_c$, where $\Delta P \ll P_c$.

{\it Effective Free Energy:}
The possibility of coexistence of SDW and TS in quasi-one-dimensional conductors
transcends microscopic descriptions based on 
standard g-ology, where SDW and TS phase boundaries 
neighbor each other but do not coexist~\cite{solyom-79}. 
Inspired by experiments~\cite{mortensen-82,lee-02b}, 
we model ${\rm (TMTSF)_2 PF_6}$ as a highly anisotropic orthorombic crystal, 
and we take the primary 
directions of the SDW vector order parameter to be the b-axis (y-direction),
and the primary direction of the TS vector order parameter to
be the c-axis (z-direction).
Furthermore, we consider the spatial variation of the 
SDW or TS order parameter to be along the a-axis (x-direction),
as a reflection of the quasi-one-dimensionality of the system. 
This simplifies the choice of the order parameters 
to be ${\bf S} ({\bf r}) \to S_b (x)$, and 
${\bf D} ({\bf r}) \to D_c (x)$, and reduces the associated 
effective field theory to one spatial dimension.
%
\begin{figure}
\begin{center}
\psfrag{P1(T)}{\scriptsize{$P_1(T)$}}
\psfrag{P2(T)}{\scriptsize{$P_2(T)$}}
\psfrag{Pc}{\scriptsize{$P_c$}}
\psfrag{Tc}{\scriptsize$T_c$}
\psfrag{P1(0)}{\scriptsize$P_1(0)$}
\psfrag{P2(0)}{\scriptsize$P_2(0)$}
\psfrag{(TMTSF)2PF6}{\bf \scriptsize$\textrm{(TMTSF)}_2\textrm{PF}_6$}
\includegraphics[width=6.4cm]{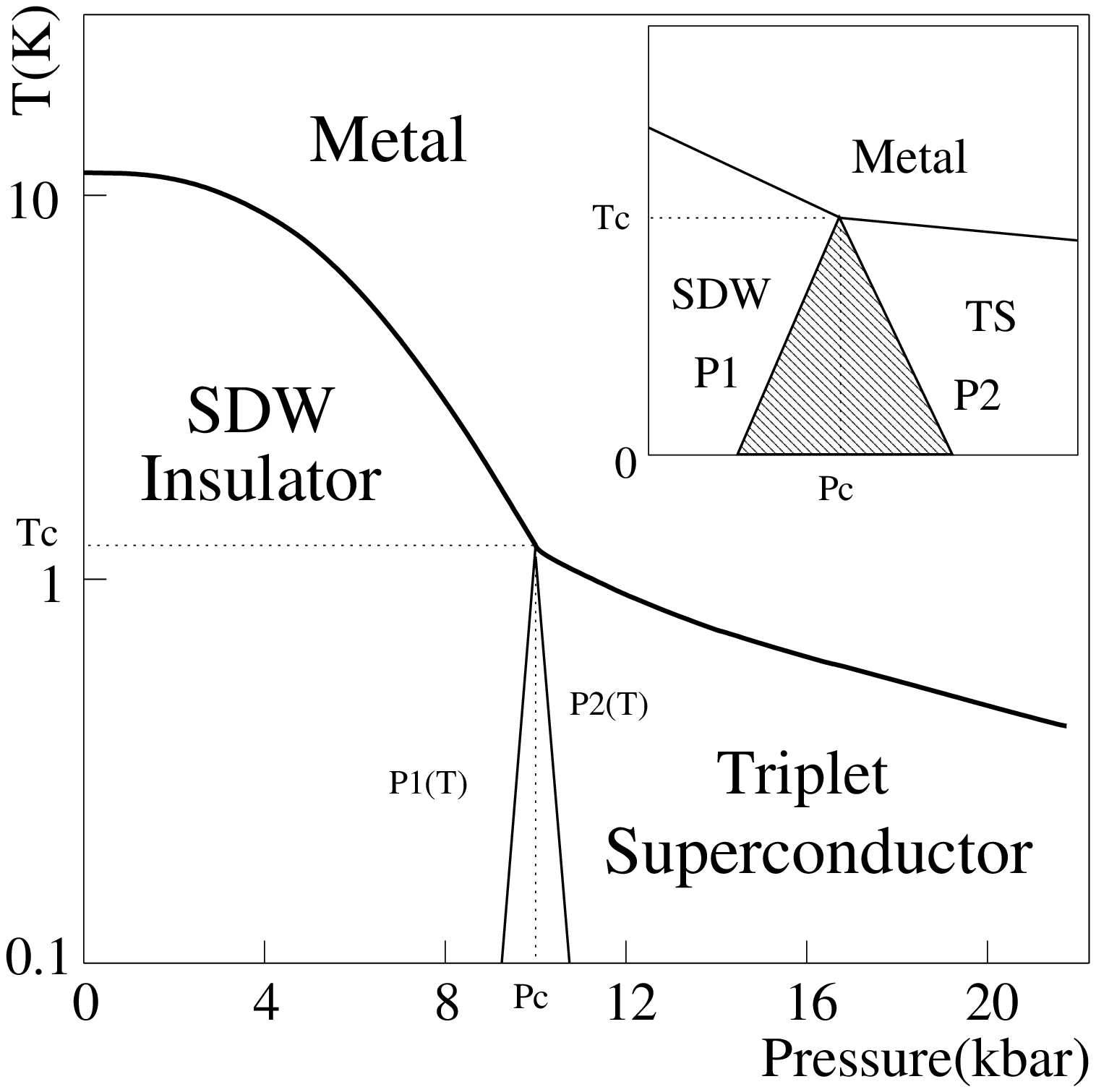}
\psfrag{lAF}{$\ell_{AF}$}
\psfrag{lTS}{$\ell_{TS}$}
\psfrag{Sb}{{\scriptsize$S_b$}}
\psfrag{Dc}{\scriptsize$D_c$}
\psfrag{Insulator}{\footnotesize{Insulator}}
\psfrag{Superconductor}{\footnotesize{Superconductor}}
\includegraphics[width=6.4cm]{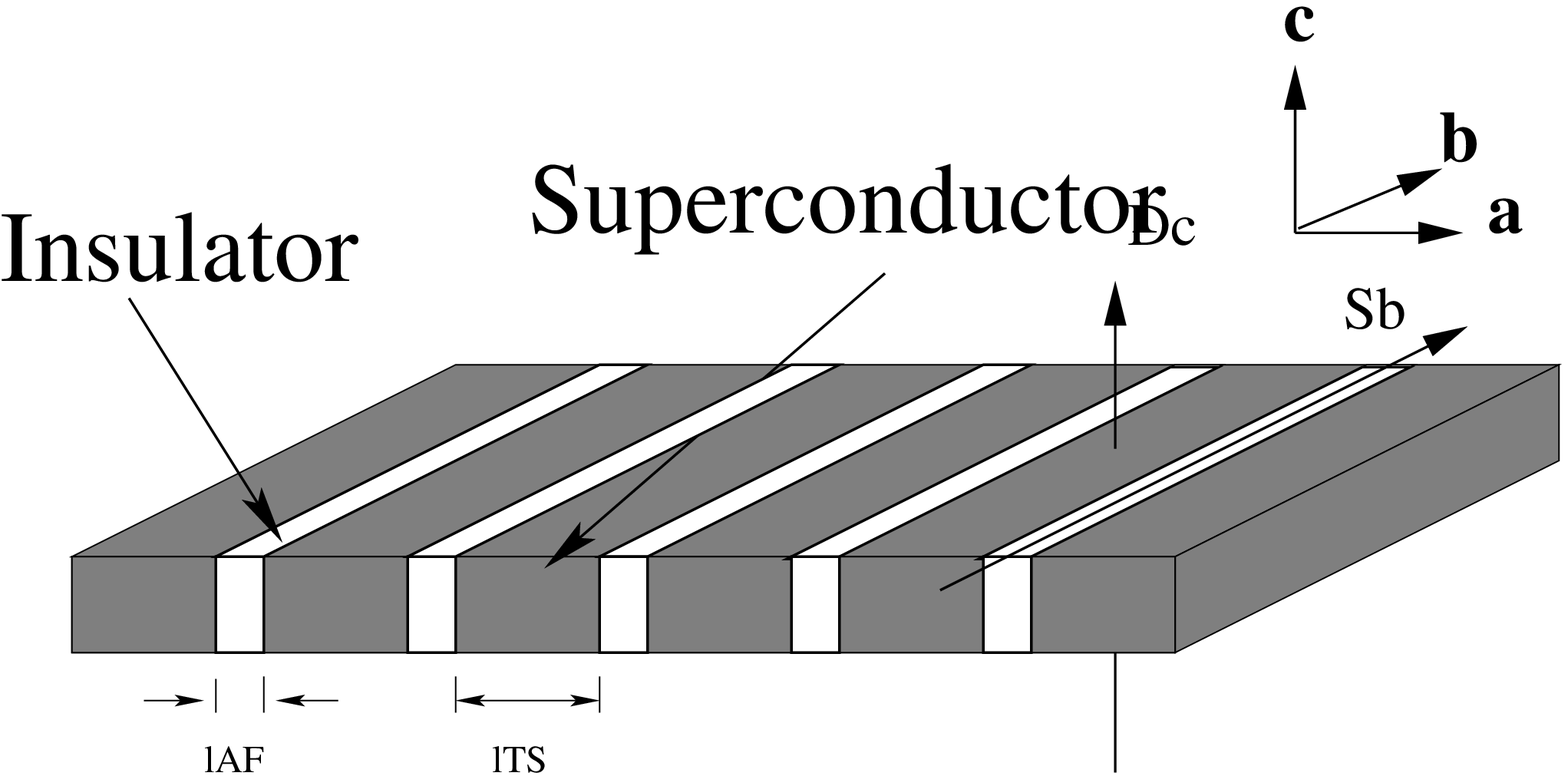}
\end{center}
\caption{ a) Phase diagram of ${\rm (TMTSF)_2PF_6}$
in a log-linear scale (from~\cite{jerome-82}), showing schematically the  
proposed SDW-TS coexistence region (inset).
b) Schematic drawing of the proposed stripe pattern for the SDW-TS 
coexistence region.}
\label{fig:phase-diagram}
\end{figure}
Thus, the generalized Ginzburg-Landau Free energy in real space can 
be written as
\begin{equation}
\label{eqn:f-tot}
\mathcal{F}_{tot}  = 
\mathcal{F}_{AF}  + 
\mathcal{F}_{TS}  + 
\mathcal{F}_{C},
\end{equation}
where $\mathcal{F}_{AF}$,  $\mathcal{F}_{TS}$, 
and $\mathcal{F}_{C}$ are the AF, TS and coupling 
contributions discussed below. 
The AF contribution is
\begin{equation}
\label{eqn:AF-FreeEnergy}
\mathcal{F}_{AF}  = 
\int_{L_{AF}}dx 
\left[ 
U_{AF} (x) + V_{AF} (x)
\right],
\end{equation}
where 
$
U_{AF}(x) = {\alpha_{AF}}{\lvert S_b(x)\rvert}^2 +
{\beta_{AF}}{ \lvert \partial_x S_b(x) \rvert }^2 +
{\gamma_{AF}}{\lvert\ S_b(x)\rvert}^4 
$
represents a typical GL Free energy density, and
$
V_{AF} = 
{\delta_{AF}}{\lvert \partial_x S_b(x) \rvert}^4 +
{\theta_{AF}}{\lvert S_b(x) \rvert}^2 {\lvert \partial_x S_b(x) \rvert}^2     
$
represents the extra terms in the expansion, which are
relevant close to $P_1 (T)$ (Fig.~1). 
The TS contribution is
\begin{equation}
\label{eqn:TS-FreeEnergy}
\mathcal{F}_{TS}  = 
\int_{L_{TS}}dx 
\left[ 
U_{TS} (x) + V_{TS} (x)
\right],
\end{equation}
where 
$
U_{TS}(x) = {\alpha_{TS}}{\lvert D_c(x)\rvert}^2 +
{\beta_{TS}}{ \lvert \partial_x D_c(x) \rvert }^2 +
{\gamma_{TS}}{\lvert\ D_c(x)\rvert}^4 
$
represents a typical GL Free energy density, and
$
V_{TS} = 
{\delta_{TS}}{\lvert \partial_x D_c(x)\rvert}^4 +
{\theta_{TS}}{\lvert D_c(x) \rvert}^2 {\lvert \partial_x D_c(x) \rvert}^2     
$
represents the extra terms in the expansion, which are 
relevant close to $P_2 (T)$ (Fig.~1). 
To describe the coexistence region 
the two order parameters must couple. 
To conform with independent Parity invariance, 
\begin{equation}
\label{eqn:Free-Energy-Coupling}
\mathcal{F}_{C} = 
\sum_{inter} 
\int_{\ell_p} dx \lambda_{bc}'
\lvert S_{b} (x)   \rvert^2
\lvert D_{c} (x)   \rvert^2,
\end{equation}
where the sum is over all interfaces between AF and 
TS, the coupling constant $\lambda_{bc}'$ is pressure 
and temperature dependent, and $\ell_p$ is the proximity length over which 
AF and TS order parameters coexist locally. This length can be 
written as $\ell_p = \ell_{p,AF} + \ell_{p,TS}$, where $\ell_{p,AF}$ 
is the AF proximity length into the TS region, and 
$\ell_{p,TS}$ is the TS proximity length into the AF region (See Fig. 2).
If $\lambda_{bc}' > 0$ it is more 
favorable for the AF and TS phases to phase-separate, however if
$\lambda_{bc}' < 0$ an inhomogeneous phase with a large number of  
interfaces is favored, and the coexistence of AF and TS is possible.

In non-triplet systems local AF order and local singlet superconductivity (SS)
can in principle coexist since AF order favors singlet correlations, 
and the proximity lengths on SS/AF systems can be small or large depending 
on the SS and AF materials~\cite{bozovic-03,bell-03}.
However, we are interested in TS and {\bf not} in SS.
In this case, it is well known
that AF order is pair breaking to triplet electron pairs~\cite{nakajima-73}, 
and it is expected that $\ell_{p, AF}$ and $\ell_{p, TS}$ are small in comparison 
to the lengths of $\ell_{TS}$ and $\ell_{AF}$ of the TS and AF stripes, 
respectively. Only when the pressure $P$ is close to the 
phase boundaries $P_1$ or $P_2$
(shown in Fig.~1) where $\ell_{TS}$ and $\ell_{AF}$ approach zero respectively, 
the proximity lengths $\ell_{p,TS}$ and $\ell_{p,AF}$ can be comparable to 
$\ell_{AF}$ and $\ell_{TS}$. Using WKB~\cite{bozovic-03} and the 
deGennes~\cite{deGennes-66} extrapolation methods
the upper bound for $\ell_p$ is 
$\ell_p (P) \le 0.1 [\ell_{AF} (P) + \ell_{TS} (P)]$, and
$\ell_{p,AF} (P) \le 0.1 \ell_{TS} (P_2)$ and
$\ell_{p,TS} (P) \le 0.1 \ell_{AF} (P_1)$. 
Thus, extrapolating AF and TS order parameters in 
the proximity region by linear functions {\it a la} deGennes~\cite{deGennes-66} 
leads to
\begin{equation}
\label{eqn:Free-Energy-Coupling-2}
\mathcal{F}_C = 
\sum_{inter} \lambda_{bc} \vert \partial_x S_b(x) \vert^2 
\vert \partial_x D_c(x) \vert^2,
\end{equation}
where $\lambda_{bc} = \lambda_{bc}' \int_0^{\ell_p} dx x^2 (\ell_p - x)^2$ 
is the new coupling constant.
This coupling describes well the inhomogeneous phase where
proximity effects between AF and TS stripes are weak, i.e.,
{\bf away} but not too far from phase boundaries $P_1$ and $P_2$.

Proximity effects will be important close to either 
phase boundary $(P_1, P_2)$ as they may lead to further coupling between like-stripes.
The inter-TS-stripe coupling (Josephson-type) 
\begin{equation}
\label{eqn:Free-Energy-TS}
\mathcal{F}_{I,TS} = \sum_{n} 
\int_{overlap} dx \eta_{TS} 
\vert D_{c, n+1}(x) - D_{c, n}(x) \vert^2
\end{equation} 
is significant for $P \approx P_2$, where the system changes from 
2D to 3D TS. 
The inter-AF-stripe coupling 
\begin{equation}
\label{eqn:Free-Energy-AF}
\mathcal{F}_{I,AF} = \sum_{n} 
\int_{overlap} dx \eta_{AF} 
\vert S_{b,n+1}(x) - S_{b,n}(x) \vert^2
\end{equation} 
is significant for $P \approx P_1$, where the system 
changes from 3D to 2D AF. 
The domain of integration for both cases above is the overlap region between two 
consecutive AF or TS stripes respectively. 

{\it Saddle Point Equations:}
To obtain the saddle point equations, we consider $\mathcal{F}_{C}$, 
$\mathcal{F}_{I,AF}$ and $\mathcal{F}_{I,TS}$ perturbatively 
and minimize $\mathcal{F}_{tot}$
with respect to $S_b (x)$ and $D_c^* (x)$. 
Variations of $\mathcal{F}_{tot}$ with respect 
to  $S_b (x)$ 
lead to the differential equation
$$
[
2\alpha_{AF} 
+ 
4 \gamma_{AF} S_b^2(x)
- 
\beta_{AF} \partial^2_x
] S_b(x) 
+
{\hat M}_{AF} S_b(x) 
= 0, 
$$
with 
$
{\hat M}_{AF} S_b(x) =
-
\delta_{AF} \partial_x \left[(\partial_x S_b(x)\right]^3 
+ 
2 \theta_{AF} S_b(x) \lvert \partial_x S_b(x) \rvert^2.
$
Variations of $\mathcal{F}_{tot}$ with
respect to $D_c^* (x)$ lead to a similar equation. 
In the case where $\lambda_{bc}' (P,T)< 0$, 
the formation of an inhomogeneous phase of alternating AF and TS stripes 
is preferred, and two additional
transition lines ($P_1$, $P_2$) emanate from $(P_c, T_c)$. 
The presence of the inter-TS and inter-AF stripe Free energies indicates 
that $D_{n+1}(x) = D_n (x + x_0)$, and $S_{n+1}(x) = S_n (x + x_0)$ (in phase
solutions), since $\eta_{AF}$ and $\eta_{TS}$ are both positive.
For such inhomogeneous phase, the boundary conditions in the presence of 
AF-TS interfaces can be chosen as in deGennes method~\cite{deGennes-66} by 
requiring that $S_b (x)\vert_{inter^+} = 0$ and 
$D_c (x)\vert_{inter^-} = 0$, where $inter^+$ and $inter^-$ denote the 
two boundaries limiting the region of locally coexisting  
$S_b(x)$ (AF) and $D_c(x)$ (TS) (See Fig.~2). 
\begin{figure}
\begin{center}
\psfrag{laf}{$\ell_{AF}$}
\psfrag{lts}{$\ell_{TS}$}
\psfrag{lpts}{$\ell_{p,TS}$}
\psfrag{lpaf}{$\ell_{p,AF}$}
\psfrag{lp}{$\ell_{p}$}
\psfrag{|S|}{$\vert S_b \vert$}
\psfrag{|D|}{$\vert D_c \vert$}
\psfrag{ltsef}{$\ell_{TS}^{eff}$}
\psfrag{lafef}{$\ell_{AF}^{eff}$}
\psfrag{+}{$+$}
\psfrag{-}{$-$}
\includegraphics[width=7.5cm]{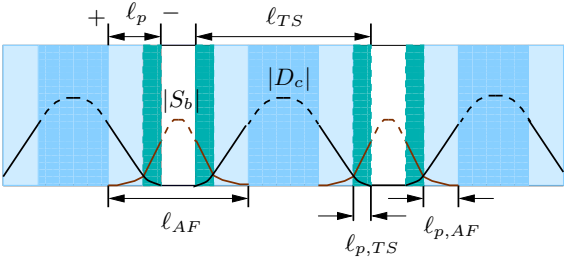}
\end{center}
\caption{Alternating AF and TS stripes with local coexistence region 
$\ell_p$, which is defined by the region between {\it interface} $^+$ and 
{\it interface} $^-$ where AF and TS order parameters vanish. 
The length $\ell_p=\ell_{p,AF} + \ell_{p,TS}$, 
where $\ell_{p,AF}$ and $\ell_{p,TS}$ are AF and TS proximity lengths.}
\end{figure}

{\it Variational Free Energy:} We analyse $\mathcal{F}_{tot}$ variationally.
We consider first the AF case and search for periodic solutions with
period $\ell_{AF}$, 
with $S_b (x)\vert_{inter^+} = 0$ at 
the AF-TS interfaces. For a given volume of the AF region, controlled
by $L_{AF}$, the Free energy associated with the AF phase becomes
the sum of $N_{AF}$ identical terms, where $N_{AF} = L_{AF}/\ell_{AF}$ 
gives the number of AF stripes. 
Generally, each term in $\mathcal{F}_{AF}$ corresponds to an 
insulating AF stripe characterized by the order parameter 
$S_b (x) = \sum_n A_n \sin (Q_n x)$, where $Q_n = 2\pi n/\ell_{AF}$.
But here we take for simplicity the variational class where
$S_b (x) = A_1 \sin (Q_1 x)$.
To simplify notation, we will just use
$A_1 \to A$ and $Q_1 \to Q$.
In this case
\begin{equation}
\label{eqn:free-af-variational}
\mathcal{F}_{AF}
=
{L_{AF}} \Big[
C_2(Q) A^2 + C_4(Q) A^4
\Big],
\end{equation}
where
$
C_2 (Q) = 
(\alpha_{AF} + \beta_{AF}Q^2)/2
$
and  
$
C_4(Q) = 
( 3 \gamma_{AF} + \theta_{AF} Q^2 + 3 \delta_{AF} Q^4 )/8.
$
The same type of analysis applies to $\mathcal{F}_{TS}$.
In the absence of magnetic field,
we assume periodic solutions of the form
$D_c (x) = \sum_n B_n \sin (K_n x)$.
As in the AF case we confine ourselves to a single 
component variational form $D_c (x) = B_1 \sin (K_1 x)$,
and use the simplifying notation $B_1 \to B$ and
$K_1 \to K$. Here $B$ can be complex, but independent
of position $x$.
All the analysis discussed for the AF (SDW) case
applies with the following change of notations:
$L_{AF} \to L_{TS}$, $A \to B$, 
$Q \to K$, $\alpha_{AF} \to \alpha_{TS}$,
$\beta_{AF} \to \beta_{TS}$, etc. Which in the TS case
leads to:
\begin{equation}
\label{eqn:free-energy-TS-variational}
\mathcal{F}_{TS}
=
{L_{TS}} \Big[
D_2(K) |B|^2 + D_4(K) |B|^4
\Big],
\end{equation}
where
$
D_2 (K) = 
(\alpha_{TS} + \beta_{TS}K^2)/2
$
and  
$
D_4(K) = 
( 3 \gamma_{TS} + \theta_{TS} K^2 + 3 \delta_{TS} K^4 )/8.
$
And the coupling Free energy is 
\begin{equation}
\label{eqn:free-energy-coupling-variational}
\mathcal{F}_{C} = N_{int} \Lambda (Q,K,\ell_p) A^2 |B|^2,
\end{equation}
where $N_{int} = 2N$ is the total number of interfaces,
$f_{int} = \Lambda (Q,K,\ell_p) A^2 |B|^2$ is the Free energy of
one interface with $\Lambda (Q,K,\ell_p) = \lambda_{bc}'  
\int_{\ell_p} dx \vert \sin(Qx) \sin[K(x_0+x)] \vert^2$,
where $x_0=\ell_{TS}-\ell_{p}$.

{\it Variational Solution:} 
Variations of $\mathcal{F}_{tot}$ with respect to 
$\phi_{AF} = A$, $\phi_{TS} = |B|$, and $q_{AF} = Q$ or $q_{TS} = K$
lead to the non-trivial solutions
\begin{equation}
\label{eqn:phi-i}
 \phi_i^2  = 
 \frac
{ 4 \beta_i \theta_i  - 24 \alpha_i \delta_i }
{ 36 \gamma_i \delta_i - \theta_i^2 },\\
\end{equation}
$\vspace{-7mm}$
\begin{equation}
\label{eqn:q-i}
 q_i^2 =
\frac 
{ \alpha_i \theta_i - 6 \beta_i \gamma_i }
{ \beta_i \theta_i - 6 \alpha_i \delta_i }.
\end{equation}
The width of each stripe then is given by
\begin{equation}
\label{eqn:l-i}
\ell_{i} = 2 \pi 
\sqrt{
\frac {\beta_i \theta_i - 6 \alpha_i \delta_i}
{ {\alpha_i} \theta_{i} - 6 \beta_{i} \gamma_{i} } 
},
\end{equation}
where $i = AF, TS$. In the case of $\lambda_{bc}' < 0$, 
the transition 
line $P_1 (T)$ corresponds to the disappearance of the pure AF (SDW) 
phase, and the transition line $P_2 (T)$ corresponds to the appearance
of the pure TS phase. 
This implies that at $P_1 (T)$ the TS stripe width is $\ell_{TS} = 0$,
while at $P_2 (T)$ the AF stripe width is $\ell_{AF} = 0$. 
Furthermore, for 
$P_1 (T) < P < P_2 (T)$, $\ell_{TS}$ ($\ell_{AF}$) increases
(decreases) with increasing pressure. In order to meet these
and the saddle point requirements, the 
parameters appearing in $\mathcal{F}_{tot}$ must behave as follows. 
We define the reduced pressure changes $\Delta P_m = [P - P_m (T)]/P_c$, 
where $m =1,2$ and the density of states $N(E_{F})$ at the Fermi energy $E_F$
to analyse the AF and TS parameters. 
For $P < P_2 (T)$, the AF parameters have the form
$\gamma_{AF} = \gamma_1 N(\epsilon_F)T_c^2 > 0$; 
$\delta_{AF} = \delta_1 N(\epsilon_F)T_c^2 > 0$;
$\eta_{AF}=\eta_1 N(\epsilon_F)T_c^2 > 0$;
$\alpha_{AF} = \alpha_1 N(\epsilon_F)T_c^2 
\vert \Delta P_2 \vert^{\varepsilon_{\alpha_{AF}}}$,
with $\alpha_1 < 0$;
$\beta_{AF} = \beta_1 N(\epsilon_F)T_c^2 \vert \Delta P_2 \vert^{\varepsilon_{\beta_{AF}}}$,
with $\beta_1 < 0$;
$\theta_{AF} = \theta_1 N(\epsilon_F)T_c^2 
\vert \Delta P_2 \vert^{\varepsilon_{\theta_{AF}}}$,
with $\theta_1 < 0$; and $36\gamma_{AF} \delta_{AF} - \theta_{AF}^2 > 0$.
For $P > P_1 (T)$,
the TS  parameters have the form
$\gamma_{TS} = \gamma_2 N(\epsilon_F)T_c^2 > 0$; 
$\delta_{TS} = \delta_2 N(\epsilon_F)T_c^2 > 0$;
$\eta_{TS}=\eta_2 N(\epsilon_F)T_c^2 > 0$;
$\alpha_{TS} = \alpha_2 N(\epsilon_F)T_c^2 \vert \Delta P_1\vert^{\varepsilon_{\alpha_{TS}}}$,
with $\alpha_2 < 0$;
$\beta_{TS} = \beta_2 N(\epsilon_F)T_c^2 \vert \Delta P_1 \vert^{\varepsilon_{\beta_{TS}}}$,
with $\beta_2 < 0$;
$\theta_{TS} = \theta_2 N(\epsilon_F)T_c^2 
\vert \Delta P_1 \vert^{\varepsilon_{\theta_{TS}}}$,
with $\theta_2 < 0$; and $36\gamma_{TS} \delta_{TS} - \theta_{TS}^2 > 0$.
Consider now, the interface terms in the region
$P_1 (T) < P < P_2 (T)$, which has the form
$\lambda_{bc}' = \lambda_0 N(\epsilon_F)T_c^2
{\rm sgn} [(P - P_1) (P - P_2)] 
\vert \Delta P_1 \vert^{\varepsilon_{\lambda_{AF}}}
\vert \Delta P_2 \vert^{\varepsilon_{\lambda_{TS}}},
$
with $\lambda_0 > 0$.
This form is required to make the interface
energy negative between $P_1 (T)$ and $P_2 (T)$.
\begin{figure}
\begin{center}
\hspace{4mm}
\psfrag{Norm. Length}{Norm. Length}
\psfrag{Norm. Amplitude}{Norm. Amplitude}
\psfrag{Pressure}{Pressure}
\psfrag{P1}{\scriptsize$P_1$}
\psfrag{P2}{\scriptsize$P_2$}
\psfrag{Pc}{\scriptsize$P_c$}
\psfrag{lAF}{$\ell_{AF}$}
\psfrag{lTS}{$\ell_{TS}$}
\psfrag{A}{$\tilde A$}
\psfrag{B}{$\tilde B$}
\includegraphics[width=7.5cm]{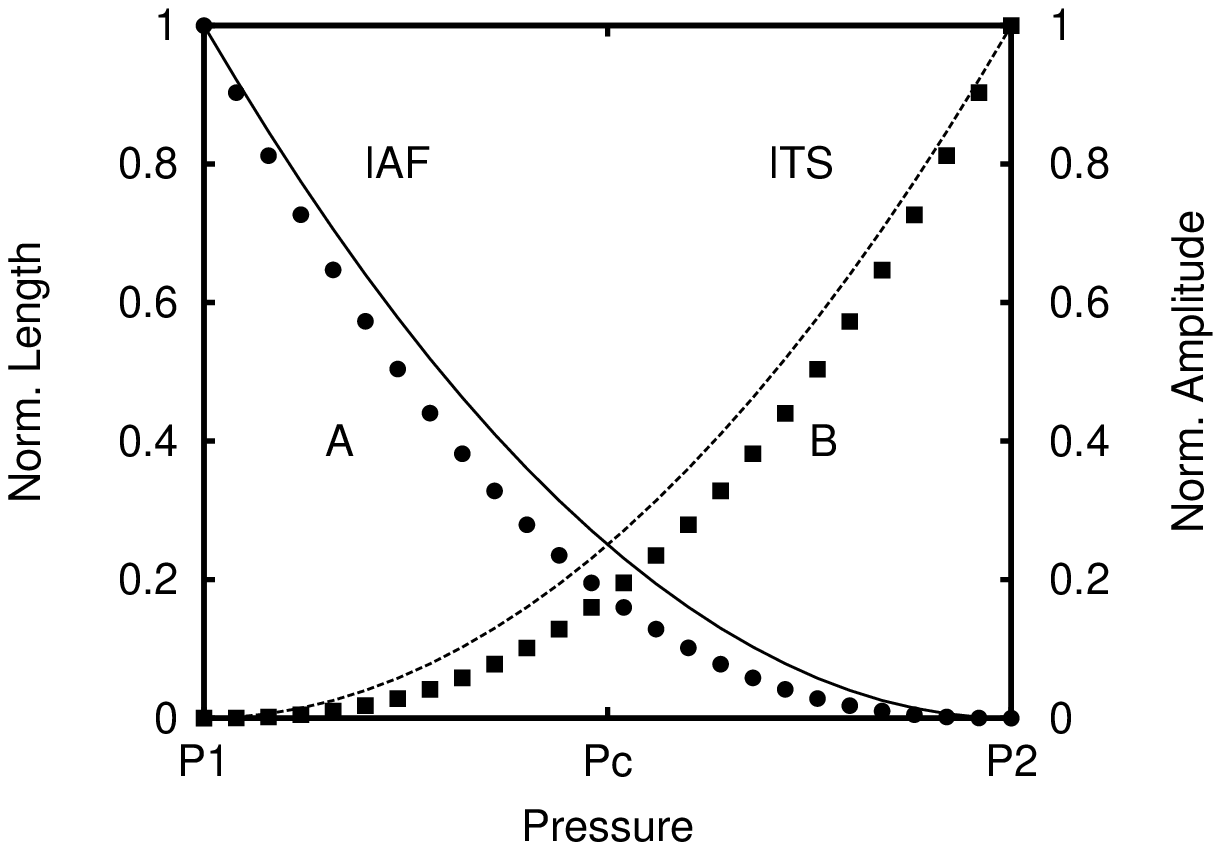}
\end{center}
\hspace{-4mm}
\psfrag{Pressure}{Pressure}
\psfrag{P1}{\scriptsize$P_1$}
\psfrag{P2}{\scriptsize$P_2$}
\psfrag{Pc}{\scriptsize$P_c$}
\psfrag{Free Energy (NefTc2)}{Free Energy ({\scriptsize$N(\epsilon_F)T_c^2$})}
\psfrag{P1'}{\scriptsize$P_1'$}
\psfrag{P2'}{\scriptsize$P_2'$}
\includegraphics[width=6.4cm]{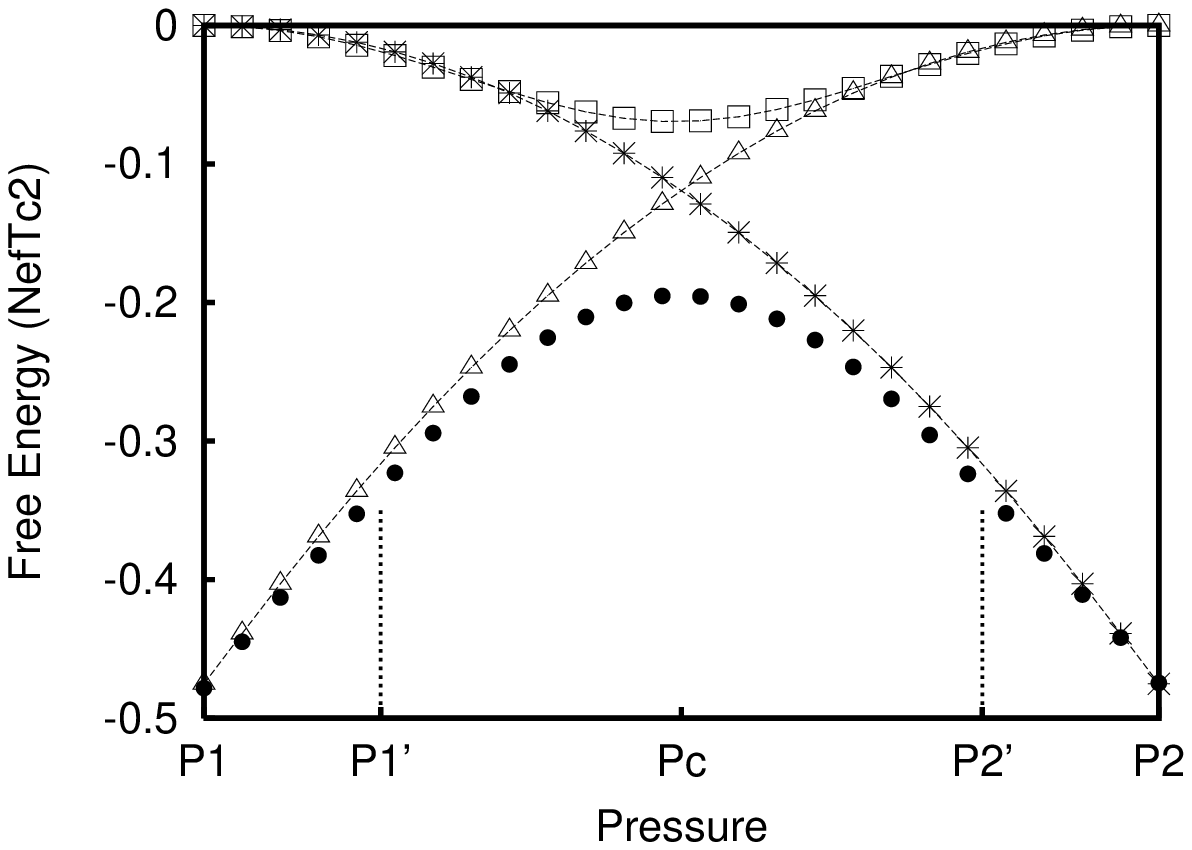}
\caption{  
a) Normalized amplitudes of the order
parameters ($\tilde A = A/ \vert A(P_1) \vert$, 
$\tilde B = B / \vert B(P_2) \vert$) 
and stripe lengths $\ell_{AF}$ and $\ell_{TS}$ normalized 
by $\ell_{AF}(P_1)$ and $\ell_{TS}(P_2)$, 
for exponents $\varepsilon_{\alpha_i} = 5.0$, 
$\varepsilon_{\beta_i} = 1.0$, $\varepsilon_{\theta_i} = 5.0$,
where $i = {AF,~TS}$ and 
dimensionless parameters: 
$\tilde\alpha_1 = \tilde\beta_1  = 1.0$,
$\tilde\gamma_1 = 0.07$,
$\tilde\delta_1 = 0.007$,
$\tilde\theta_1 = -0.003$;
and 
$\tilde\alpha_2 = \tilde\beta_2  = 1.0$,
$\tilde\gamma_2 = 0.06$,
$\tilde\delta_2 = 0.007$,
$\tilde\theta_2 = -0.002$.
b) Free energies for the coexistence and the
pure AF and TS phases for the same parameters
of a) and $\ell_p=0.05(\ell_{AF}+\ell_{TS})$, 
$\varepsilon_{\lambda_i} = 1.0$, 
$\tilde\lambda_0 = -0.003$, 
and $\tilde\eta_1=\tilde\eta_2= 2.9\times 10^{-6}$. 
Solid dots $\to$
$\mathcal{F}_{tot}$; stars $\to$ $L f_{TS}$; 
triangles $\to$ $L f_{AF}$;
squares $\to$ $\mathcal{F}_{C}$. 
$P_1'$ indicates change from 3D to 2D AF, and 
$P_2'$ denotes change from 2D to 3D TS.
}
\label{fig:length-order-parameter}
\end{figure}

{\it Phase Boundaries:}
Next, we focus only on the analysis of $\ell_{AF}$ 
and $\mathcal{F}_{tot}$ in the vicinity of
$P_2(T)$. We note in passing that the analysis of
$\ell_{TS}$ and $\mathcal{F}_{tot}$ in the vicinity of
$P_1(T)$ is entirely analogous.
Under these considerations, 
as $P \to P_2 (T)$ the size of the AF stripes is given by
$
\ell_{AF} \approx W_1
\vert \Delta P_2 \vert^{(\varepsilon_{\alpha_{AF}}
 - 
\varepsilon_{\beta_{AF}} )/2}, 
$
where
$W_1 = 2\pi (\alpha_1 \delta_1 / \beta_1 \gamma_1)^{1/2}$,
when $\varepsilon_{\beta_{AF}} + \varepsilon_{\theta_{AF}} 
> \varepsilon_{\alpha_{AF}}$,
and $\varepsilon_{\alpha_{AF}} + \varepsilon_{\theta_{AF}} 
> \varepsilon_{\beta_{AF}}$.
The requirement that $\ell_{AF} \to 0$ as $P \to P_2 (T)$ forces
$\varepsilon_{\alpha_{AF}} >  \varepsilon_{\beta_{AF}}$. 
Since the number of AF and TS stripes are the same $N_{AF} = N_{TS} = N$,
where $N = L /(\ell_{AF}(P)+ \ell_{TS}(P)-\ell_p(P))$, the number
of interfaces is $N_{int} = 2N$. Therefore, the four contributions
to $\mathcal{F}_{tot}$ in the 
coexistence region are
$\mathcal{F}_{AF} = L_{AF}f_{AF}$, 
$\mathcal{F}_{TS} = L_{TS} f_{TS}$,
$\mathcal{F}_{C} = 2 N f_{int}$, 
$\mathcal{F}_{I,TS}= 2 N f_{I,TS}$ 
with 
$L_{AF}(p)= N \ell_{AF}(P)$, $L_{TS}= N \ell_{TS}(P)$.
As $P \to P_2$, the AF stripe length
$\ell_{AF}(P) \to 0$, while the TS stripe length  
$\ell_{TS}(P) \to \ell_{TS}(P_2 (T))$.
Let us analyse the behavior of $\mathcal{F}_{tot}$ near $P_2 (T)$
term by term.   
The AF part $\mathcal{F}_{AF}$ is the product of
$L_{AF}(P) \approx L \ell_{AF}(P)/\ell_{TS}(P_2)$,  
where 
$\ell_{AF}(P) = Const. \vert \Delta P_2 \vert^{(\varepsilon_{\alpha_{AF}}
-\varepsilon_{\beta_{AF}})/2}$,
and $f_{AF} = - Const. \vert \Delta P_2  \vert^{2\varepsilon_{\beta_{AF}}}$,
which leads to 
$\mathcal{F}_{AF} =  
- Const. \vert \Delta P_2 \vert^{(\varepsilon_{\alpha_{AF}} 
+ 3\varepsilon_{\beta_{AF}})/2}$.
The TS part  $\mathcal{F}_{TS}$ is the product of
$L_{TS}(P)  \approx  L [1-\ell_{AF}(P)/\ell_{TS}(P_2)]$ 
and $f_{TS}(P) \approx f_{TS}(P_2) < 0$, thus 
$\mathcal{F}_{TS} \approx \mathcal{F}_{TS} (P_2) 
+ L Const.\vert \Delta P_2 \vert^{(\varepsilon_{\alpha_{AF}}
-\varepsilon_{\beta_{AF}})/2}$, where
$\mathcal{F}_{TS} (P_2) = L f_{TS} (P_2)$ is the Free energy
of the pure triplet phase. 
Thus, for $P < P_2$,
$\mathcal{F}_{TS}$ is increased with respect to the pure phase
$\mathcal{F}_{TS} (P_2)$.
Furthermore, 
$\mathcal{F}_{C} = - Const. \vert P_2 - P \vert^{ \varepsilon_{\lambda_{AF}} 
+ \varepsilon_{\beta_{AF}} }$ is negative, 
and $\mathcal{F}_{I,TS}$ is positive but proportional to
a higher power of $\Delta P_2$. 
$\mathcal{F}_{AF}$ can be neglected in the vicinity of $P_2$
because it depends on a higher power of $\vert \Delta P_2 \vert$ than that of 
$\mathcal{F}_{TS}$ or $\mathcal{F}_{C}$, when 
$\varepsilon_{\alpha_{AF}} > \varepsilon_{\beta_{AF}}$. 
For $\mathcal{F}_{tot}$ to be lower than that
of the pure TS phase $\mathcal{F}_{TS} (P)$ it is necessary that
the negative interface Free energy $\mathcal{F}_{C}$ dominates. 
This imposes the following requirement 
$\varepsilon_{\alpha_{AF}} \ge 3\varepsilon_{\beta_{AF}} 
+ 2\varepsilon_{\lambda_{AF}}$. 
If $\varepsilon_{\lambda_{AF}} + \varepsilon_{\beta_{AF}} > 1$ 
the phase transition
at $P_2$ is continuous.
If $\varepsilon_{\lambda_{AF}} + \varepsilon_{\beta_{AF}} \le 1$ 
the phase transition
at $P_2$ is discontinuous. 
Similar analysis close to $P_1$ leads to a continuous
transition when 
$\varepsilon_{\lambda_{TS}} + \varepsilon_{\beta_{TS}} > 1$ 
and to a discontinuous transition
when  $\varepsilon_{\lambda_{TS}} + \varepsilon_{\beta_{TS}} \le 1$. 
Thus, the point $(P_c, T_c)$ can be bicritical, tricritical or
tetracritical. In Fig.~3, we show the behavior of the various contributions to
$\mathcal{F}_{tot}$ for the case where the transitions are
continuous at $P_1$ and $P_2$, and $(P_c, T_c)$ is tetracritical.
Dimensionless ``volume'' parameters are defined as
$\tilde\alpha_i = \tilde\beta_i =  \rho_i^{1/2}$,  
$\tilde\gamma_i = \gamma_i \sigma_i^{-1/2}$,
$\tilde\delta_i = \delta_i \sigma_i^{3/2}$,
$\tilde\theta_i = \theta_i \sigma_i^{1/2}$,
``surface'' ones are defined as 
$\tilde\lambda_0 = \lambda_0 \sigma_1^{-1/4} \sigma_2^{-1/4}(\ell_p(P_1) \sigma_1^{1/2})^5$,
$\tilde\eta_i = \eta_i \sigma_i^{-1/2}(\ell_p(P_1) \sigma_1^{1/2})^3$,
where $\rho_i = \alpha_i \beta_i $,
$\sigma_i = \alpha_i/\beta_i$, and $i = 1, 2$.

{\it Final Comments:} 
A more realistic description of the system should include
variations of SDW and TS order parameters along the y and z (transverse) 
directions, and be 
highly anisotropic but truly three dimensional. 
This is important for
a renormalization group (RG) analysis and the determination 
of critical exponents in the case of continuous transitions. 
             
{\it Summary:} We have proposed the possibility of coexistence
of antiferromagnetism and triplet superconductivity in 
the phase diagram of ${\rm (TMTSF)_2 PF_6}$. This intermediate phase
is proposed to be inhomogeneous and to consist of alternating 
insulating AF and TS stripes~\cite{demler-04}
Two additional transition lines appear in a narrow 
range of pressures around $P_c$ separating the coexistence region
from the pure AF and pure TS phases. We estimate the maximum 
pressure range to be $\Delta P/P_c \approx 10\% $ at $T = 0$.

{\it Acknowledgements:}
We thank NSF for support (DMR-0304380), and  
Ivan Bozovic for discussions.

\end{document}